\documentclass[a4paper,20pt]{article}
    \usepackage[T1]{fontenc}
    \usepackage{graphicx}
    \usepackage{epsfig}
    \usepackage{amsmath}
    \usepackage{amsfonts}
    \renewcommand{\abstract}{}
\begin{document}
\makeatletter
\renewcommand{\@oddhead}{\textit{Advances in Astronomy and Space Physics} \hfil \textit{A.V. Tugay}}
\renewcommand{\@evenfoot}{\hfil \thepage \hfil}
\renewcommand{\@oddfoot}{\hfil \thepage \hfil}
\fontsize{11}{11} \selectfont

\title{Bright X-ray galaxies in SDSS filaments}
\author{\textsl{A.\,V.~Tugay}}
\date{}
\maketitle
\begin{center} {\small 
Taras Shevchenko National University of Kyiv, Glushkova ave., 4, 03127, Kyiv, Ukraine\\
tugay@anatoliy@gmail.com}
\end{center}

\begin{abstract}
Eighteen bright X-ray emitting galaxies were found in nearby filaments within SDSS region. Basic X-ray spectral parameters were estimated for these galaxies using power law model with photoelectric absorption. A close pair of X-ray galaxies was found. 

\end{abstract}

\section*{Introduction}
\indent \indent The large-scale structure (LSS) of observable Universe contains clusters and groups of galaxies combined into filaments. The problem of (galaxy) filaments detection in redshift space is quite complicated, so the methods of its solving are usually tested on syntetic LSS obtained from N-body simulations. Smith et al. (2012, \cite{smith12}) found 53 real filaments for Sloan Digital Sky Survey galaxies at redshift z<0.13. In this paper the filaments were defined as elongating structures containing enough groups and clusters to build minimal spanning tree in redshift space. Such definition lead that filaments cover only a little fragment of SDSS redshift-space volume available for LSS studying. New methods for filament detection should be developed to describe the structure of large and continuous space volumes. An attempt of developing and application of such new metod (single galaxies are used in this method instead of groups and clusters) to SDSS volume will be the subject of our future work. Now we will consider only 53 filaments established in \cite{smith12}.

 The goal of this work is to describe X-ray emission of galaxies in filaments which can be compared in the future with other high energy sources in the large-scale structures of the Universe (isolated galaxies, clusters, walls). In \cite{tugay11} it was shown that outside of Local Supercluster it is possible to detect in X-ray band only the galaxies with active nuclei (AGN), so the filaments must contain some bright X-ray emitting AGN's. X-ray galaxies in the filaments are established in \cite{smith12} and their spectra are considered here. 

\section*{Samples}

\indent \indent A sample of 5021 X-ray galaxies from \cite{tugay12} was used for the search of filament galaxies. These galaxies were selected from 2XMMi catalogue \cite{watson09} of X-ray sources detected on XMM-Newton observatory. 2092 of them lie in the main SDSS region of sky ($110^o<\alpha <250^o, -10^o<\delta <70^o$). Only the brightest X-ray sources  are appropriate for spectral fitting. After the inspection of preliminary spectra of 2XMMi sources in Vizier database, the minimal limiting value of X-ray flux $F_{X min}=3.7\cdot 10^{-13} erg/cm^2$ was picked. It is assumed here that X-ray spectra can be built only for galaxies with $F_X>F_{X min}$. 978 from all 5021 X-ray galaxies satisfy this condition. 

The second condition for galaxy selection was that the galaxy should inside one of 53 SDSS filaments established in \cite{smith12}. These filaments were found with multiscale probability methods in which the size of structures in redshift space $\Delta z=0.005$ is assumed. In the present work filaments from \cite{smith12} were visually inspected on the sky distribution of galaxies in slices with $\Delta z=0.023$ that corresponds to the size of void (100 Mpc). 53 filaments from \cite{smith12} occupy only five slices of 100 Mpc. The example of the slice is shown on Fig. 1. Numbers of galaxies in slices are presented in Table 1. Among 335 bright X-ray galaxies in SDSS region only 18 were found in the mentioned filaments. During the selection some galaxies inside or very close to X-ray clusters were detected. Although that clusters can be considered as the part of a filament, such galaxies were excluded to avoid the problem of division of radiation from galaxy and from intergalactic hot gas. To estimate the upper bound of possible number of X-ray galaxies in filaments, there was made the assumption that all galaxies outside clusters are located in filaments. General characteristics of the obtained galaxies are presented in Table 2. 

\begin{figure}[!h]
\begin{minipage}[t]{.99\linewidth}
\centering
\epsfig{file = 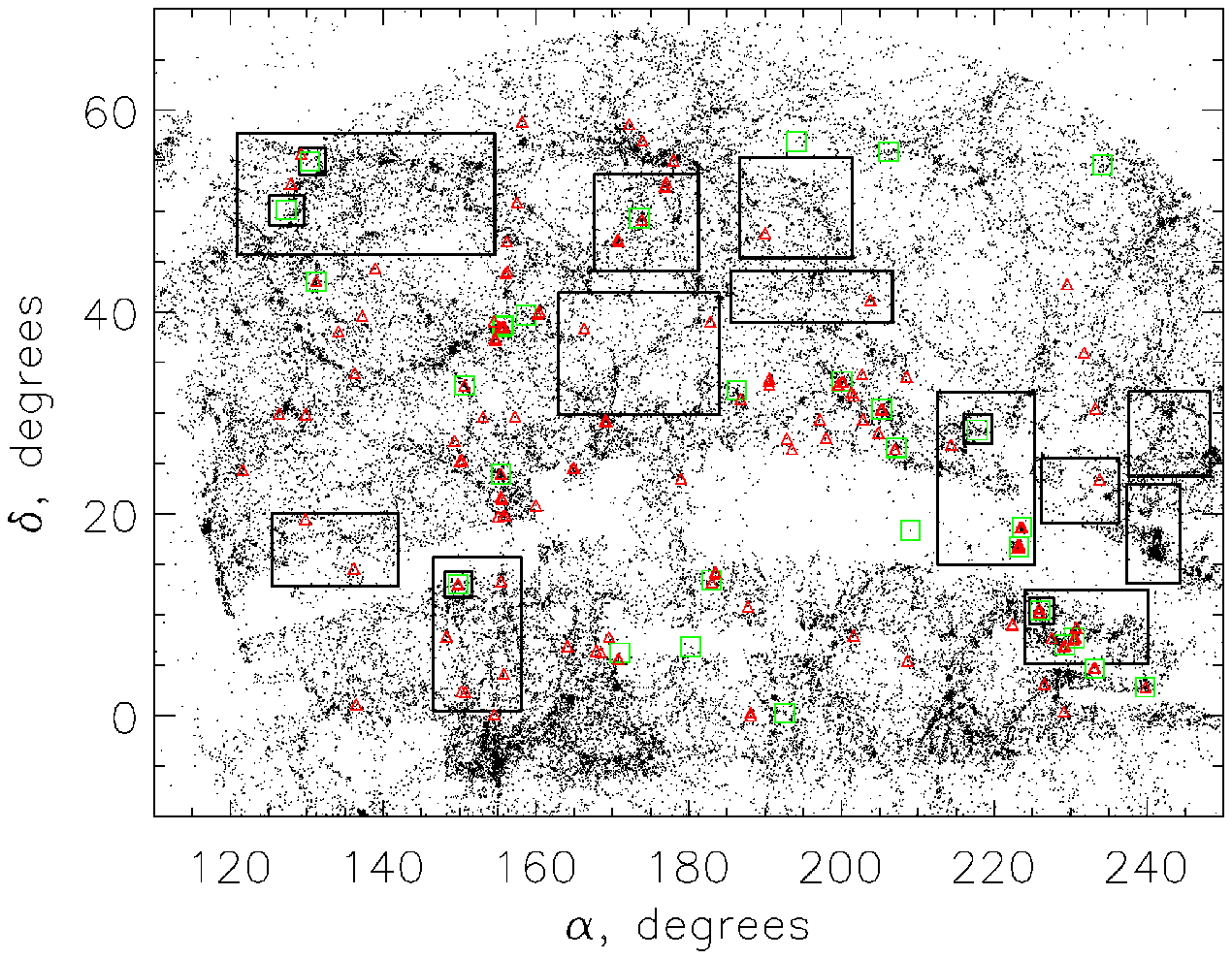, width = 0.99\linewidth}
\caption{Sky distribution of SDSS galaxies with radial velocities between 11000 and 18000 km/s. Triangles - X-ray galaxies with $F_X<F_{X min}$, squares - bright X-ray galaxies with $F_X>F_{X min}$, large rectangles - filaments from \cite{smith12}, double squares - galaxies in filaments which are considered in this work.}\label{fig7}
\end{minipage}
\end{figure}

\begin{table}
 \centering
 \caption{Number of X-ray galaxies in SDSS by distances.}\label{tab1}
 \vspace*{1ex}
 \begin{tabular}{|c|c|c|c|c|}
  \hline
  Radial velocity, km/s & Filaments in \cite{smith12}& X-ray galaxies in SDSS & Notable Objects & Bright X-ray galaxies \\
  \hline
  0-4000      &   & 207 & Local Supercluster & 79 \\
  4000-11000  &  5 & 214 & Coma Supercluster & 60 \\
  11000-18000 & 12 & 196 & & 31 \\
  18000-25000 & 21 & 212 & Sloan Great Wall & 37 \\
  25000-32000 & 10 & 148 & & 29 \\
  32000-39000 & 5 & 113 & & 12 \\
  39000-46000 &  & 109 & & 10 \\
  46000-53000 &  &  70 & & 11 \\
  53000-60000 &  &  73 & & 7 \\
  Distant &    & 555 & & 59\\
  Total & 53 & 2092 & & 335 \\
 \hline 
 \end{tabular}
\end{table}

\begin{table}
 \centering
 \caption{General properties of bright X-ray galaxies.}\label{tab1}
 \vspace*{1ex}
 \begin{tabular}{|l|c|c|c|c|}
  \hline
  Object name & Coordinate code & SIMBAD Type & $V_{3K}$, km/s & X-ray flux, $10^{-12} erg/cm^2$\\
  \hline
2XMMi J082912.8+500652  &	 0829+5007 &	 Seyfert 1  &	 13141 &	1.91$\pm$0.11\\
2MASX J08413787+5455069 &	 0842+5455 &	 Seyfert 1  &	 13512 &	3.83$\pm$0.14\\
2MASX J09590662+1301351 &	 0959+1301 &	 Galaxy    &	 11296 &	1.25$\pm$0.03\\
2MASX J09591475+1259161 &	 0959+1259 &	 Galaxy    &	 10619 &	1.39$\pm$0.03\\
2MASS J10003549+0524285 &	 1001+0524 &	 QSO  &	 23905 &	0.99$\pm$0.04\\
Mrk 176                 &	 1133+5257 &	 Seyfert 1  &	  8325 &	0.62$\pm$0.02\\
NGC 3921                &	 1151+5505 &	 Interacting Galaxies  &	  6017 &	0.42$\pm$0.01\\
2MASX J12034921+0205575 &	 1204+0206 &	 Seyfert 1  &	 24714 &	1.09$\pm$0.11\\
2E 2620                 &	 1214+1403 &	 Seyfert 1  &	 24581 &	10.97$\pm$0.02\\
NGC 4686                &	 1247+5432 &	 Galaxy in Group  &	  5181 &	1.13$\pm$0.03\\
MCG+07-28-006           &	 1324+4318 &	 Seyfert 2  &	  8374 &	1.12$\pm$0.04\\
2XMM J134245.8+403913   &	 1343+4039 &	 X-ray source    &	 26889 &	0.59$\pm$0.03\\
2MASS J14134834+4400141 &	 1414+4400 &	 Seyfert 1  &	 26842 &	0.74$\pm$0.02\\
NGC 5548                &	 1418+2508 &	 Seyfert 1  &	  5358 &	74.44$\pm$0.13\\
Mrk 684                 &	 1431+2817 &	 Seyfert 1  &	 13949 &	7.54$\pm$0.06\\
87GB 150115.3+102756    &	 1501+1028 &	 AGN  &	 28668 &	0.55$\pm$0.05\\
Mrk 841                 &	 1504+1026 &	 Seyfert 1  &	 11126 &	35.02$\pm$0.15\\
2MASX J15113367+0545479 &	 1512+0546 &	 AGN  &	 25065 &	0.43$\pm$0.01\\
 \hline 
 \end{tabular}
\end{table}

\begin{table}
 \centering
 \caption{Spectral properties of bright X-ray galaxies.}\label{tab1}
 \vspace*{1ex}
 \begin{tabular}{|l|c|c|c|c|c|c|}
  \hline
 Target & Galactic $N_H, 10^{22}cm^{-2}$ & $N_H, 10^{22}cm^{-2}$ & Photon index & $\chi ^2$/d.o.f. & Source\\
  \hline
0829+5007 	 &0.046 	& 0.05 	                                                 &2.64$\pm$0.08 	                                        &102.4/101	 &\cite{dewangan08}\\
0842+5455 	 &0.038 	& 0.31$\pm$0.02 	                                 &1.56$\pm$0.06 	                                        &109/135	 &This work\\
0959+1301 	 &0.032 	& 12.7$\pm$2.5 	                                         &1.34$\pm$0.29 	                                        &15.8/16	 &This work\\
0959+1259 	 &0.032 	& 0.77 $\pm$ 0.05 	                                 &1.88 $\pm$ 0.08 	                                        &206/267	 &\cite{lamassa09}\\
1001+0524 	 &0.027 	& <0.01 	                                         &2.26$\pm$0.11 	                                        &35/28	         &This work\\
1133+5257 	 &0.011 	& 12.7$\pm$2.5 	                                         &1.34$\pm$0.29 	                                        &11.4/11	 &This work\\
1151+5505 	 &0.010 	& 1.93$\pm$0.20 	                                 &1.55$\pm$0.10 	                                        &93/86	         &\cite{nolan04}\\
1214+1403 	 &0.030 	& 0.043$\pm$0.006 	                                 &2.07$\pm$0.01 	                                        &896/746	 &This work\\
1247+5432 	 &0.015 	&0.024$\pm$0.001	                                 &2.00$\pm$0.05	                                                &119/61	         & \cite{vasudevan13}\\
1324+4318 	 &0.015 	& 0.06$\tiny \begin{matrix} +0.34\\ -0.04\end{matrix}$ 	 &2.74$\tiny \begin{matrix} +2.40\\ -0.67\end{matrix}$ 	        &141/68	         &\cite{lamassa09}\\
1343+4039 	 &0.008 	& 3.8$\pm$1.2 	                                         &1.25$\pm$0.37 	                                        &10.6/11	 &This work\\
1414+4400 	 &0.009 	& 16.7$\tiny \begin{matrix} +4.5\\ -5.6\end{matrix}$  	 &1.55$\tiny \begin{matrix} +0.24\\ -0.23\end{matrix}$ 	        &63/75	         &\cite{inoue07}\\
1418+2508 	 &0.017 	& 0.02 	                                                 &1.65 	                                                 	&                &\cite{vasudevan09}\\
1431+2817 	 &0.016 	&<0.01	                                                 &2.45$\pm$0.02 	                                        &400/286	 &This work\\
1501+1028 	 &0.024 	&<0.01	                                                 &1.48$\pm$0.01 	                                        &1022/770	 &This work\\
1504+1026 	 &0.024 	& 0.02 	                                                 &1.75$\pm$0.05 	                                   	&                &\cite{cerruti11}\\
1512+0546 	 &0.036 	& 0.08$\pm$0.02 	                                 &2.22$\pm$0.19 	                                        & 62/58	         &This work\\
 \hline 
 \end{tabular}
\end{table}

\section*{Spectral fitting}
\indent \indent Cross-correlation of XMM and SDSS sources was performed in \cite{pineau11} but there was no detailed study of single galaxies. 
Basic spectral characteristics of bright filament galaxies are presented in Table 3. 
Power law model with photoelectric absorbtion was used in this analysis. Galactic absorbtion was taken from \cite{willingale13}. Spectral parameters for some galaxies, when possible, were taken from literature. Nine galaxies has no spectral fitting in previous works, so their spectra were built in this work. That galaxies were mentioned previously only in AGN catalogs such as \cite{veron10} and their cross-correlation with XMM sources in \cite{pineau11}. 

There are some notes for few chosen galaxies.

2MASX J12034921+0205575 This Seyfert 1 galaxy has five references in SIMBAD database including \cite{pineau11}, \cite{veron10} and 3 older optical surveys. 2XMMi database contains preliminary spectrum for this galaxy but full archive of observation data files is not available for dowloading. 

2MASX J09590662+1301351 and 2MASX J09591475+1259161. This pair of bright X-ray galaxies has angular distance of 3 arcmin that corresponds to spatial distance 140 kpc. They can interact or be physically bound.

\section*{Conclusions}
\indent \indent Eighteen bright X-ray galaxies in SDSS filaments were found. The most common type of such galaxies is Seyfert 1. Spectral parameters for nine galaxies were estimated. Since the number of X-ray sources in filaments is very little, X-ray observations can't be used for the detection of new filaments. So the procedure of describing LSS in high energy band should include detailed description of filaments on the basis of optical galaxy surveys, statistical study of X-ray sources in large homogeneous volume and comparison of parameters of X-ray extragalactic sources inside and outside filaments. Such study is going to be performed in our future works.

\section*{Acknowledgement}
\indent \indent Thanks to the referee for very useful advises in emproving this work and propositions of its continuation. The author also acknowledges to A.Vasylenko for his help with spectral fitting.

\end{document}